\documentclass[review,12pt]{elsarticle}

\usepackage{amssymb}

\usepackage{graphicx}
\usepackage{xcolor}
\usepackage{hyperref}
\hypersetup{
    colorlinks=true,
    linkcolor=blue,
    citecolor=blue,
    urlcolor=blue,
}
\usepackage{float}
\usepackage{caption}
\usepackage{subcaption}

\biboptions{numbers,sort&compress}

\journal{Elsevier}

\begin{document}

\begin{frontmatter}



\title{Arrhenius.jl: A Differentiable Combustion Simulation Package}


\author[mit]{Weiqi Ji \corref{cor1}}
\ead{weiqiji@mit.edu}

\author[thu]{Xingyu Su}

\author[weichai]{Bin Pang}

\author[stan]{Sean Joseph Cassady}
\author[stan]{Alison M. Ferris}

\author[weichai]{Yujuan Li}
\author[thu]{Zhuyin Ren}
\author[stan]{Ronald Hanson}

\author[mit]{Sili Deng \corref{cor1}}
\ead{silideng@mit.edu}
\cortext[cor1]{Corresponding author}

\affiliation[mit]{organization={Department of Mechanical Engineering},
            addressline={Massachusetts Institute of Technology}, 
            city={Cambridge},
            state={MA},
            postcode={02139},
            country={USA}}

\affiliation[thu]{organization={Center for Combustion Energy},
            addressline={Tsinghua University}, 
            city={Haidian},
            state={Beijing},
            postcode={100084},
            country={China}}

\affiliation[weichai]{organization={Weichai Power Co. Ltd.},
            city={Weifang},
            state={Shandong},
            country={China}}

\affiliation[stan]{organization={High Temperature Gasdynamics Laboratory, Department of Mechanical Engineering},
            addressline={Stanford University}, 
            city={Stanford},
            state={CA},
            postcode={94305},
            country={USA}}



\begin{abstract}
Combustion kinetic modeling is an integral part of combustion simulation, and extensive studies have been devoted to developing both high fidelity and computationally affordable models. Despite these efforts, modeling combustion kinetics is still challenging due to the demand for expert knowledge and optimization against experiments, as well as the lack of understanding of the associated uncertainties. Therefore, data-driven approaches that enable efficient discovery and calibration of kinetic models have received much attention in recent years, the core of which is the optimization based on big data. Differentiable programming is a promising approach for learning kinetic models from data by efficiently computing the gradient of objective functions to model parameters. However, it is often challenging to implement differentiable programming in practice. Therefore, it is still not available in widely utilized combustion simulation packages such as CHEMKIN and Cantera. Here, we present a differentiable combustion simulation package leveraging the eco-system in Julia, including DifferentialEquations.jl for solving differential equations, ForwardDiff.jl for auto-differentiation, and Flux.jl for incorporating neural network models into combustion simulations and optimizing neural network models using the state-of-the-art deep learning optimizers. We demonstrate the benefits of differentiable programming in efficient and accurate gradient computations, with applications in uncertainty quantification, kinetic model reduction, data assimilation, and model discovery.
\end{abstract}


\end{frontmatter}


\section{Introduction}
The optimization of model parameters plays a critical role in the development of chemical kinetic simulation tools. While heuristic optimization methods, such as genetic algorithms, have been widely employed in combustion modeling, optimization algorithms based on stochastic gradient descent (SGD)  have seldom been exploited. Meanwhile, SGD has shown promise in nonconvex optimization for complex nonlinear models, and SGD has played a central role in driving the boom of deep learning in the last decade \cite{bengio2017deep}.

Existing optimization techniques in combustion modeling can be categorized into heuristic algorithms and response surface techniques. Heuristic algorithms \cite{rein2006application, bertolino2021evolutionary} usually perform well with less than 100 parameters and small datasets. In the case of the genetic algorithm, for example, the computational cost scales with the number of parameters and the number of samples in the dataset. As a result, modern deep learning working in big data regimes and deep neural network models with tens of thousands of parameters seldom employ heuristic optimization. Response surface techniques \cite{sheen2011method} alleviate the intensive computational cost in evaluating kinetic models by building a function approximation that maps the model parameter space to model predictions. Similar to heuristic algorithms, response surface techniques are also limited to low-dimensional model parameter space \cite{ji2018shared} and small datasets, as the cost of building a response surface scales with the dimension of parameter space and the number of quantities of interest. Meanwhile, recent development in sensor techniques and experiment automation \cite{tao2019kinetic} have significantly increased the efficiency of experimental data generation and driven combustion research into the big data regime. Combined with SGD, the optimization of complex chemical models using these new datasets becomes feasible. Furthermore, various techniques have been developed in conjunction with SGD to increase the generalization performance of the optimized model. For instance, a modern deep learning optimizer not only focuses on minimizing the loss functions but also regularizes the model to increase the extrapolation capability. Generalization to different conditions and tasks is an important feature for classical physics-based chemical models, and generalization allows us to develop a chemical model based on canonical combustion experiments that also works reasonably well in simulating practical combustion systems. Therefore, SGD is not only more efficient but also more generalizable compared to heuristic optimization algorithms.

One of the major obstacles for exploiting SGD in combustion modeling is the lack of software ecosystems that can efficiently and accurately compute the gradient of simulation output to model parameters. For instance, the finite difference method (often termed the brute-force method) usually suffers both computational inefficiency, as the cost scales with the number of parameters, and inaccuracy due to truncation error. Conversely, stochastic gradient descent based on auto-differentiation (AD) has shown both efficiency and accuracy in the training of large-scale deep neural network models \cite{bengio2017deep}. Many open-source AD packages have been developed in the last decade, including TensorFlow \cite{abadi2016tensorflow} and Jax \cite{jax2018github} backed by Google, PyTorch \cite{paszke2019pytorch} backed by Facebook, ForwardDiff.jl \cite{revels2016forward} and Zygote.jl \cite{innes2018don} in Julia. To this end, we introduce the AD-powered differentiable combustion simulation package Arrhenius.jl \cite{jiarrheniusgithub}, which incorporates combustion physics models into AD ecosystems in Julia to facilitate the study of differentiable combustion modeling.

This paper is structured as follows: we shall first introduce the package Arrhenius.jl in Sec. \ref{sec:arrhenius} and the formulas for various gradient calculations in Sec. \ref{sec:grad}. We then present the application of Arrhenius.jl to uncertainty quantification, kinetic model reduction, data assimilation, and model discovery in Sec. \ref{sec:results}. Finally, we draw conclusions in Sec. \ref{sec:conclusion}.

\section{Arrhenius.jl} \label{sec:arrhenius}
Arrhenius.jl is built with the programming language of Julia to leverage the rich ecosystems of auto-differentiation and differential equation solvers. Arrhenius.jl does two types of differentiable programming: (i) it can differentiate elemental computational blocks. For example, it can differentiate the reaction source term with respect to kinetic and thermodynamic parameters as well as species concentrations. (ii) It can differentiate the entire simulator in various ways, such as solving the continuous sensitivity equations \cite{ji2019evolution} as done in CHEMKIN \cite{kee1989chemkin} and Cantera \cite{goodwin2009cantera} and in adjoint methods \cite{rackauckas2018comparison, rackauckas2017differentialequations}. The first type of differentiation is usually the basis of the second type of higher-level differentiation. Arrhenius.jl offers the core functionality of combustion simulations in native Julia programming, such that users can conveniently build applications on top of Arrhenius.jl and exploit various approaches to do high-level differentiation.

\begin{figure}[H]
    \centering
    \includegraphics[width=1.0\linewidth]{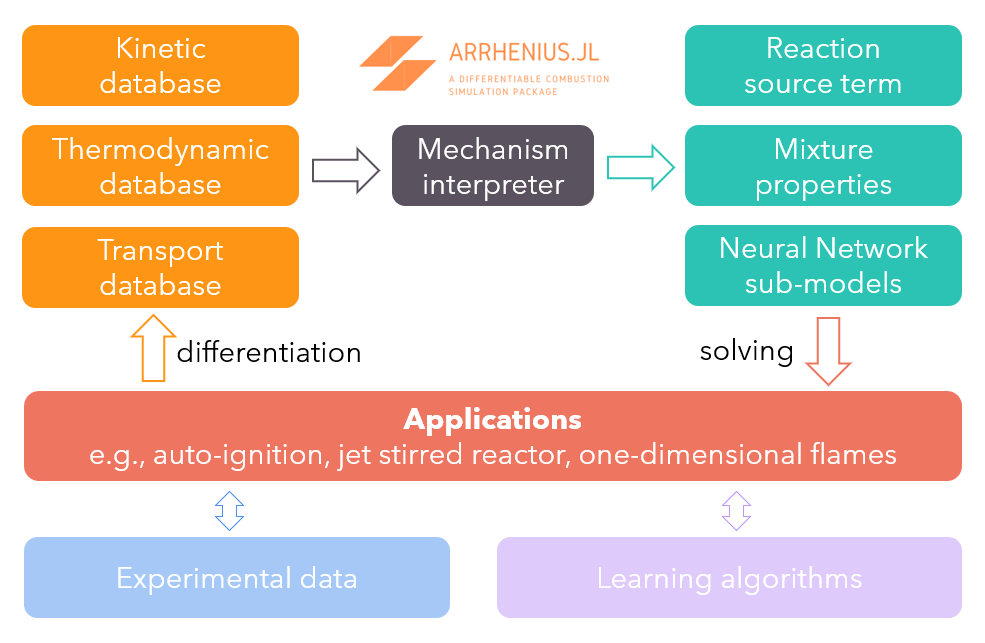}
    \caption{Schematic showing the structure of the Arrhenius.jl package.}
    \label{fig:schem}
\end{figure}

Figure \ref{fig:schem} shows a schematic of the structure of the Arrhenius.jl package. Arrhenius.jl reads in the chemical mechanism files in YAML format maintained by the Cantera community; the chemical mechanism files contain the kinetic model, thermodynamic, and transport databases. The core functionality of Arrhenius.jl is to compute the reaction source terms and mixture properties, such as heat capacities, enthalpies, entropies, Gibbs free energies, etc. In addition, Arrhenius.jl offers flexible interfaces for users to define neural network models as submodels and augment them with existing physical models. For example, one can use a neural network submodel to represent unknown reaction pathways and exploit various scientific machine learning methods to train the neural network models, such as neural ordinary differential equations \cite{chen2018neural,rackauckas2020universal} and physics-informed neural network models \cite{raissi2019physics,ji2020stiff}. One can then implement the governing equations for different applications with these core functionalities and solve the governing equations using classical numerical methods or neural network-based solvers, such as physics-informed neural networks \cite{raissi2019physics}. Arrhenius.jl provides solvers for canonical combustion problems, such as simulating the auto-ignition in constant volume/pressure reactors and oxidation in jet-stirred reactors.

In contrast to legacy combustion simulation packages, Arrhenius.jl can not only provide predictions given the physical models, but also optimize model parameters given experimental measurements. By efficiently and accurately evaluating the gradient of the solution outputs to the model parameters, experimental data can be incorporated into the simulation pipeline to enable data-driven modeling with deep learning algorithms.

\section{Gradient Calculation} \label{sec:grad}

In this section, we briefly discuss how Arrhenius.jl can enhance various approaches for computing the gradient (sensitivity) in jet-stirred reactors, shock tubes, and laminar flame experiments. In general, we deal with two kinds of gradient calculations, i.e., steady-state solutions and transient solutions.

Examples of steady-state solutions are the modeling of species profiles in jet-stirred reactors and steady laminar flames. Without loss of generality, we can write down the governing equations in vector form as
\begin{equation}
    F( \phi (\alpha); \alpha) = 0,
    \label{eq:steady_govern}
\end{equation}
where $F$ corresponds to the residual vector, $\phi$ corresponds to the solution vector, and $\alpha$ corresponds to model parameters, such as the kinetic, thermodynamic, and transport parameters. By differentiating Eq. \ref{eq:steady_govern} with respect to the $\alpha$, we obtain matrix equations for the gradients.
\begin{equation}
    \frac{\partial F}{\partial \phi} \frac{\partial \phi}{\partial \alpha} + 
    \frac{\partial F}{\partial \alpha} = 0,
    \label{eq:steady_grad}
\end{equation}
where $\frac{\partial \phi}{\partial \alpha}$ are the Jacobian matrices of the solution vector with respect to model parameters. Normally, for optimization, we have a scalar loss function defined as $L = G(\phi)$. The gradient of the loss function with respect to model parameters can be readily achieved via
\begin{equation}
    \frac{\partial L}{\partial \alpha} = \frac{\partial L}{\partial \phi} 
            \frac{\partial \phi}{\partial \alpha}.
    \label{eq:steady_loss}
\end{equation}

With Arrhenius.jl, we can leverage AD to compute the two Jacobian matrices of $\frac{\partial F}{\partial \phi}$ and $\frac{\partial F}{\partial \alpha}$. Unlike packages for calculating the analytical Jacobian, efficient computation of the Jacobian can be achieved without developing the analytical form. In addition, general AD enables us to differentiate over neural network submodels, which is difficult to implement using analytical approaches. If the solution variables are discretized in a computational domain, e.g., the one-dimensional freely propagating flame, one can readily leverage multi-threading to evaluate the two Jacobians without complex code re-factorization using parallel computing.

Examples of transient solutions are auto-ignition and fuel pyrolysis in shock tubes. The governing equations are in the general form of
\begin{equation}
    \frac{d\phi}{dt} = F( \phi, t; \alpha).
    \label{eq:trans_govern}
\end{equation}

A natural approach to compute the gradient $W = \frac{\partial \phi}{\partial \alpha}$ is solving the governing equations of $W$, i.e.,
\begin{equation}
    \frac{dW}{dt} =  \frac{\partial F}{\partial \phi} W + \frac{\partial F}{\partial \alpha}.
    \label{eq:trans_grad}
\end{equation}

In addition to solving Eq. \ref{eq:trans_grad}, Arrhenius.jl leverages various adjoint sensitivity algorithms provided in DifferentialEquations.jl. For comprehensive comparisons of various algorithms in computing the gradient of solutions variables for transient solutions, readers shall consult \cite{rackauckas2018comparison}.

In general, calculating the gradient for transient solutions involving stiff chemical kinetic models is expensive, as computation time usually scales with both the number of species and the number of parameters. Meanwhile, global combustion behaviors, such as ignition delay times (IDTs), are usually more experimentally accessible compared to measurements of concentration profiles. Recent work \cite{gururajan2019direct, ji2019evolution, lemke2019adjoint, almohammadi2021tangent} has substantially advanced the algorithms for computing the gradient of IDT. This work employs the sensBVP method \cite{gururajan2019direct}, which converts the initial value problem (IVP) to a boundary value problem (BVP) by treating the temperature at ignition as a boundary condition and the IDT as a free variable to solve. In other words, the problem is converted from a transient problem to a steady-state problem, similar to the solution of a one-dimensional freely propagating flame.

\section{Results and Discussion} \label{sec:results}

We now present four case studies using Arrhenius.jl for uncertainty quantification, kinetic model reduction, data assimilation, and model discovery.

\subsection{Application to Uncertainty Quantification} \label{sec:uq}
We first apply Arrhenius.jl to computing the active subspace for quantifying kinetic uncertainties. Active subspace \cite{constantine2014active} is a parametric dimension reduction approach that identifies the low-dimensional subspace of high-dimensional uncertain parameters via the singular value decomposition of the expected parameter's gradient. It has been applied to evaluate combustion model uncertainties in zero/one-dimensional simulations \cite{ji2018shared, vohra2019active, su2021uncertainty}, as well as in turbulent combustion simulations \cite{ji2019quantifying, wang2020quantification,wang2021active}. The active subspace corresponds to the directions along which the simulation results change significantly, while the simulation results are almost unchanged in the kernel space. Uncertainty quantification in combustion simulations usually suffers from the curse of dimensionality, i.e., the computational cost exponentially increases with the number of model parameters. With the active subspace, one can conduct various uncertainty quantification tasks within the low-dimensional subspace. The active subspace methodology is detailed in \cite{constantine2014active} and its key idea is illustrated in Eq. \ref{eq:active_subspace}:
\begin{equation}
    \mathbf{C} = \int \nabla f(\mathbf{x}) {\nabla f(\mathbf{x})}^T 
        \pi_\mathbf{x}(\mathbf{x}) d\mathbf{x} = \mathbf{W} \Lambda \mathbf{W}^T,
    \label{eq:active_subspace}
\end{equation}
where $\mathbf{x} \in \mathcal{R}^d$ represents the samples from the parameter space, $\pi_\mathbf{x}$ is the distribution of the model parameters, $f(\mathbf{x})$ refers to the quantity of interest and $\nabla f(\mathbf{x})$ corresponds to the gradient with respect to the model parameters. The unitary matrix $\mathbf{W}$ consists of the $d$ eigenvectors $\mathbf{w}_1, \mathbf{w}_2, ..., \mathbf{w}_d$ and $\mathbf{\Lambda}$ is a diagonal matrix whose components are the eigenvalues $\lambda_1, \lambda_2, ..., \lambda_d$, sorted in descending order. If there is a gap in the eigenvalues, meaning $\lambda_r >> \lambda_{r+1}$, then the function $f(\mathbf{x})$ varies mostly along the first $r$ eigenvectors and is almost constant along the rest of the eigenvectors. The first e eigenvectors are selected as active directions, i.e., $\mathbf{S} = [\mathbf{w}_1, \mathbf{w}_2, ..., \mathbf{w}_r]$. The matrix $\mathbf{C}$ is usually approximated by Monte Carlo sampling and the number of samples required is suggested to be $m = \alpha \beta log(d)$. The constant $\alpha$ is the over-sampling factor and is recommended to be between 2 and 10, and $\beta$ is the largest dimension of the subspace allowed in subsequent utilization of active subspace. The major challenge for exploiting active subspace in combustion modeling is how to efficiently compute the gradient $\nabla f(\mathbf{x})$. With Arrhenius.jl, we now can efficiently and accurately compute the gradients and thus make the active subspace approach more widely accessible in combustion modeling.

We used the chemical models of GRI3.0 and LLNL's detailed $n$-heptane model (Version 3.1) \cite{mehl2011kinetic} for demonstrations to compute eigenvalue spectra. The ignition delay time (IDT) is specified as our quantity of interest. The thermodynamic conditions for both cases are 40 atm, 1200 K, and stoichiometric fuel/air mixture. The sensBVP approach is employed to compute the gradient of IDT to the pre-exponential factors $A$ of the reactions in the kinetic model. An independent log-uniform distribution is assumed for each reaction, and $ln(A/A_0) \sim \mathbf{U}[-0.5, 0.5]$, in which $A_0$ denotes the nominal value of the pre-exponential factor A, and $\mathbf{U}$ denotes a uniform distribution. The number of samples drawn from the parameter space is set as $m = 20 * log(d)$ for methane, $10 * log(d)$ for $n$-heptane, and $d$ is the number of reactions.

\begin{figure}[H]
    \centering
    \includegraphics[width=1.0\textwidth]{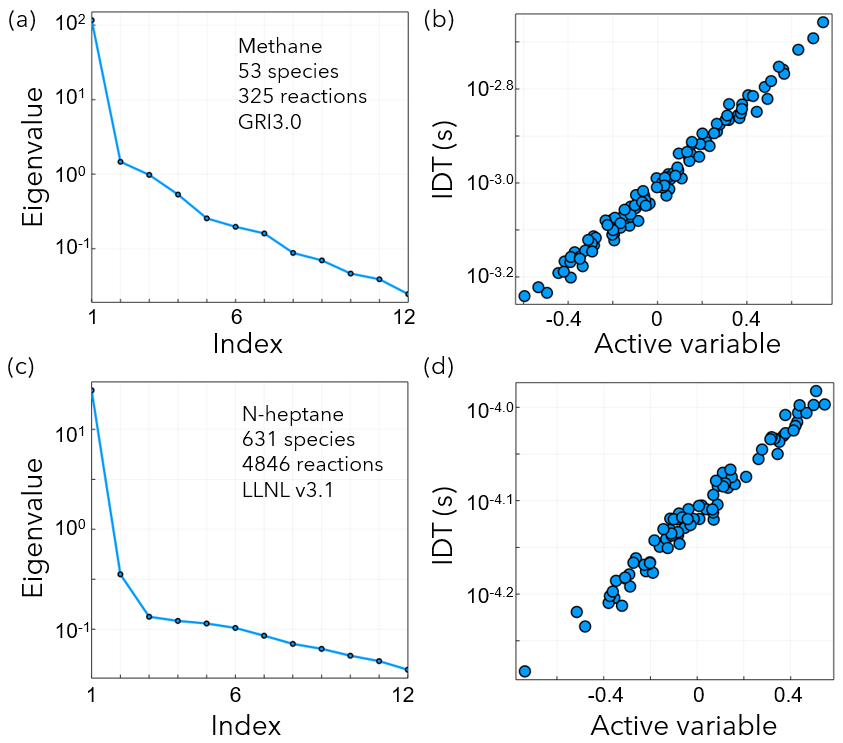}
    \caption{(a, c) Eigenvalue spectra  and (b, d) summary plots for the models of GRI3.0 and LLNL v3.1 for $n$-heptane. The thermodynamic conditions for both cases are under 40 atm, 1200 K, and stoichiometric fuel/air mixture. The uncertainties considered are the uncertainties in the pre-exponential factor $A$. An independent log-uniform distribution is assumed for all of the reactions, and $ln(A/A_0) \sim \mathbf{U}[-0.5, 0.5]$. The number of samples drawn from the parameter space is set as $20 * log(nr)$ for methane, $10 * log(nr)$ for $n$-heptane, and $nr$ is the number of reactions.}
    \label{fig:activesubspace}
\end{figure}

In Figs. \ref{fig:activesubspace}a and c, one-dimensional active subspace can be identified for both cases, since the first eigenvalue is larger than the second by two orders of magnitude. The identified active subspace is shown in the summary plot, where all of the samples are plotted against the first active direction defined by the leading eigenvector. The summary plots in Figs. \ref{fig:activesubspace}b and d further show that the variations in IDT can be well captured by the first active direction, for all samples are distributed along a one-dimensional curve. Those results demonstrate the capability of Arrhenius.jl in effectively identifying the active subspace. With the one-dimensional active subspace identified, one can exploit it for global sensitivity analysis, forward and inverse uncertainty quantification, and optimization.

As far as the computational cost is concerned, our approach obtained the active subspace within one minute on a laptop for GRI3.0, while the finite difference approach took about an hour. In addition, the model for n-heptane consists of 4846 reactions, and 4846 is the highest dimension that has ever been achieved in uncertainty quantification for chemical models in literature, to the best knowledge of the authors.

\subsection{Application to Kinetic Model Reduction} \label{deepreduction}
Most skeletal chemical models are obtained by removing unimportant species and the associated reactions from the master model, leaving the kinetic parameters of the remaining pathways unchanged after the reduction. There has also been increasing interest in optimizing the kinetic parameters in overly reduced reaction models to compensate for the error introduced by over-reduction \cite{mittal2017multi, kelly2021toward}; this has the potential to produce a smaller model with higher fidelity than the ones obtained via traditional reduction methods. The reduction-optimization approach proceeds by sampling from target conditions and then optimizing the reduced model to achieve similar predictions as the master mechanism under all sampled conditions.

Furthermore, our recent work \cite{ji2021autonomous} has shown that a chemical reaction network is equivalent to a neural network with a single hidden layer. Similarly, solving ordinary differential equations (ODEs) of reaction network models is equivalent to solving infinite-depth deep residual networks \cite{chen2018neural}. Consequently, chemical model reduction is in analogy to the compression of deep neural networks (DNNs). The compression of DNNs is essential for the wide application of deep learning to reduce the size and hence the computational cost of DNNs. Deep Compression \cite{han2015deep} is one of the most famous compression techniques, which utilizes a multi-stage pipeline for compression. It first applies weight pruning to remove unimportant connections in the neural network, similar to the pathway removal in skeletal mechanism reduction for chemical models. It then applies quantization and retrains the compressed network. Inspired by Deep Compression, we develop Deep Reduction, a two-stage reduction scheme, with the first step being the conventional skeletal reduction and the second step being fine-tuning the reduced model using an SGD optimizer.

The Deep Reduction approach is demonstrated in reducing and optimizing two chemical models for natural gas and $n$-heptane, the master models of which are the GRI3.0 mechanism \cite{smith1999gri} and the Nordin1998 mechanism \cite{nordin1998numerical}, respectively. As previously discussed, overly reduced models using classical reduction approaches with a large threshold or intuition are first obtained, with the number of species in GRI3.0 reduced from 53 to 23, and that in Nordin1998 reduced from 41 to 34. The reduction is targeted for simulating natural gas engines and diesel engines with the commercial software of Converge. For GRI3.0, we first removed the following species using an iterative reduction involving DRG \cite{lu2006applicability}, DRGEP \cite{pepiot2005systematic}, PFA \cite{sun2010path}, and sensitivity analysis: {C, CH$_3$OH, C$_2$H, C$_2$H$_2$, HCCO, CH$_2$CO, HCCOH,
NH, NH$_2$, NH$_3$, N$_2$O, HNO, CN, HCN, H$_2$CN, HCNN, HCNO, HOCN, HNCO, NCO, Ar, and CH$_2$CHO}. We then further removed the species of CH$_3$CHO, NO$_2$, NO, NNH, N, C$_2$H$_3$, and CH$_2$OH, CH by removing NO-chemistry for Converge has a built-in NO module and following the reduction of DRM19 \cite{kazakov1994reduced}. The overly reduced model is denoted as SK23 with 23 species. For Nordin1998, the following species were removed based on the reduction of GRI3.0 above: C$_3$H$_5$, C$_3$H$_4$, C$_2$H$_6$, CH$_4$O$_2$, CH$_3$O$_2$, CH$_3$O, and C$_2$H$_2$. The overly reduced model is denoted as SK34 with 34 species. Note that the way to produce an overly reduced model can be regarded as a hyper-parameters subject to explore. This work focuses on the demonstration of the optimization algorithms, and we leave the optimal choice of removed species to future studies. It should be noted that one can also optimize an existing empirical semi-global reaction model against a detailed model without consulting the skeletal mechanism reduction.

\begin{figure}[!h]
    \centering
    \includegraphics[width=1.0\textwidth]{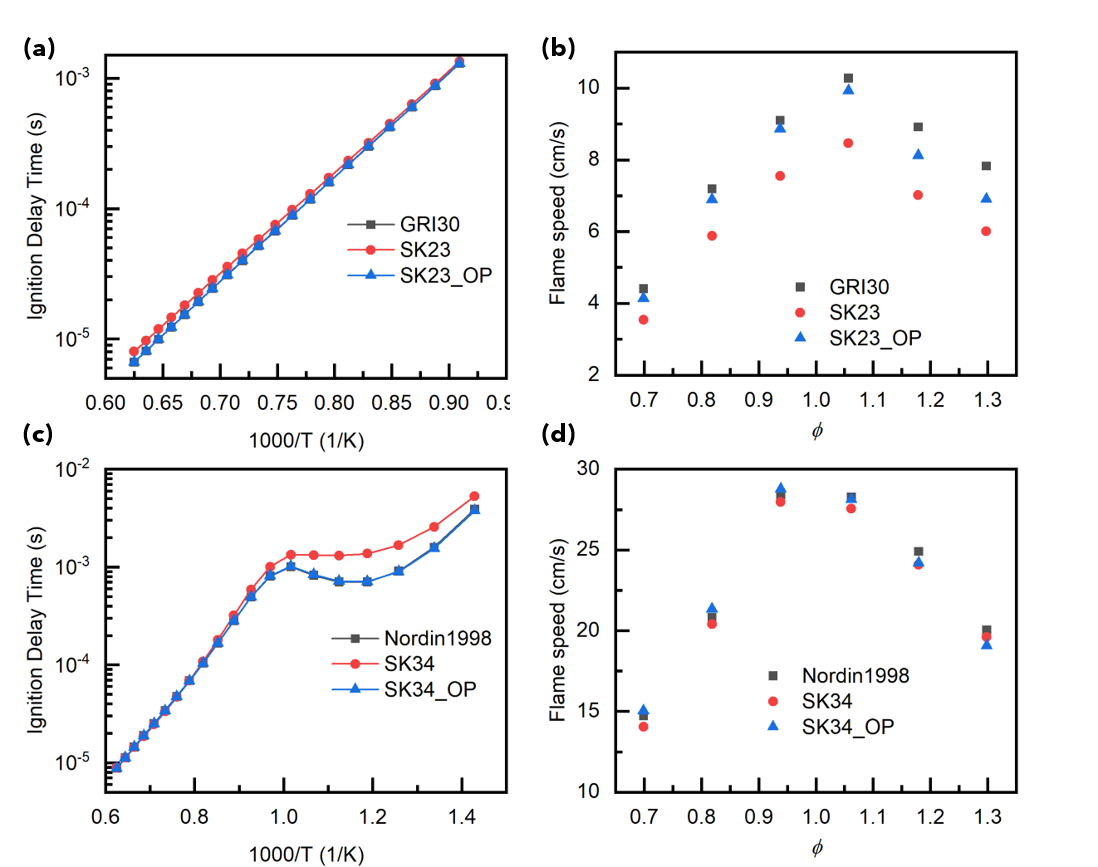}
    \caption{Predicted ignition delay times and flame speeds: (a-b) the mixture of natural gas/air using the master mechanism GRI3.0, skeletal mechanism SK23 and optimized SK23\_OP. The ignition delay time was simulated using the fuel composition of CH$_4$ : C$_2$H$_4$ : C$_3$H$_8 = 0.85:0.1:0.05$ by volume, pressure of 40 atm, equivalence ratio of 0.9. The flame speed was simulated at 40 atm and 300 K, and (c-d) the mixture of $n$-heptane/air using the master mechanism Nordin1998, skeletal mechanism SK34 and optimized SK34\_OP. The ignition delay time was simulated at pressure of 40 atm, equivalence ratio of 1.2. The flame speed was simulated at 40 atm and 500 K.}
    \label{fig:deepreduction_combined}
\end{figure}

The kinetic parameters of these overly reduced models were subsequently optimized to retain the predictability of the master models. The ignition delay times and laminar flame speeds (SLs) are utilized as performance metrics to validate the overly reduced and optimized models against the corresponding master models. As shown in Fig. \ref{fig:deepreduction_combined}, for GRI3.0, SK23 overpredicts IDT at high temperatures and underpredicts SL at all equivalence ratios. For Nordin1998, SK34 significantly overpredicts the IDT at low temperatures, especially within the negative temperature coefficient region, while the flame speeds are hardly affected by the reduction.

The optimization of the kinetic parameters of the over-reduced models was conducted on all three Arrhenius parameters, namely, $A$, $b$, $E_a$. Although both IDT and SL could be selected as targets for optimization, we only utilized the IDT for its relatively lower computational cost than SL. Moreover, the top ten reactions selected based on the sensitivity analysis for the SL were excluded from the optimization, such that the optimization will not change these key reactions for SL. We then randomly sampled 500 initial conditions covering a wide range of mixture compositions and thermodynamic states for training. For GRI3.0, the range of the pressure is 1-60 atm, that of the initial temperatures is 1100-2000 K, that of the equivalence ratios is 0.5-1.8, and the fuel composition is set as CH$_4$ : C$_2$H$_4$ : C$_3$H$_8$ $= 0.85:0.1:0.05$ by volume. Similarly, for Nordin1998, the ranges of the pressure, initial temperature, and equivalence ratio are 1-60 atm, 850-1800 K, and 0.5-1.5, respectively.

The datasets were split into training and validation datasets with a ratio of 70:30. During each parameter update, one case was randomly sampled to evaluate its IDT, and this process can be viewed as mini-batching with the batch size of one. Instead of optimizing the Arrhenius parameters directly, we optimized the relative changes of Arrhenius parameters compared to their nominal values, i.e.,
\begin{equation}
    p = [ln(A/A_0), b-b_0, Ea - Ea_0],
    \label{eq:pdef}
\end{equation}
where the subscript $_0$ refers to the base model. The units of $Ea$ are specified as $cal/mol$ as we tried to minimize the changes in $Ea$. However, if one wants to ensure the change in $Ea$ is relatively comparable to the change in $A$, one may specify the unit of $Ea$ as $kcal/mol$, as a change of 2 $kcal/mol$ in $Ea$ is close to change of $e$ times in $A$, such that the changes in $A$ and $Ea$ will be balanced and avoid stiffness in the parameter space.

\begin{figure}[!h]
    \centering
    \includegraphics[width=1.0\textwidth]{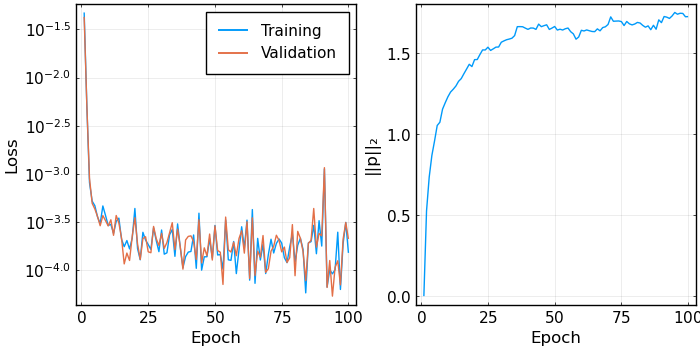}
    \caption{Training history of loss functions, $L_2$-norm of model parameters. Regularization settings: weight-decay of 1e-4 with a learning rate of 1e-3.}
    \label{fig:nordin1998_loss}
\end{figure}

The loss function was defined as the mean square error (MSE) between the predicted IDTs in the logarithmic scale using the reduced model and the master model:
\begin{equation} \label{eq:loss}
    Loss = MSE\left(log(IDT^{sk}), log(IDT^{master}). \right)
\end{equation}
The gradients of IDT to kinetic parameters were computed using the sensBVP method proposed in \cite{gururajan2019direct}. The Adam \cite{kingma2014adam} optimizer with the default learning rate of 0.001 was adopted. Weight decaying and early stopping were employed to regularize the parameters, such that the optimization prefers kinetic parameters that are close to their original values. Figure \ref{fig:nordin1998_loss} shows the training history of the loss function and the $L_2$-norm of the model parameters for the Nordin1998 model. We trained the reduced model for 100 epochs and stopped the training when the loss function as well as the model parameters reached a plateau.
 
 The performance of the optimized skeletal mechanisms is also shown in Fig. \ref{fig:deepreduction_combined}. For the IDT, as targeted in the optimization, the optimized models agree with the master models very well for both two fuels. One interesting observation is that the optimized models also work well for SL, although SL is not targeted for optimization. This could be attributed to several reasons. For GRI3.0, the optimization compensates for the errors in the high-temperature chemistry seen in the predicted IDT, and those re-calibrated high-temperature chemical pathways lead to accurate predictions of SL. For Nordin1998, the skeletal models already accurately predict the IDT at high temperatures as well as the SL, and the optimization does not degrade the prediction of SL thanks to the regularization. It is worth noting that the optimized SK23 still underpredicts SL at fuel-rich conditions, potentially because some of the sensitive reactions for fuel-rich conditions are not fixed during optimization; further refinement of the fixed reactions is therefore suggested. Furthermore, the optimization is computationally efficient. Qualitatively speaking, previously employed genetic algorithms have to be performed on clusters \cite{kelly2021toward}, while the current work was performed on an ordinary workstation within an hour. In summary, these two case studies demonstrate the ability of Arrhenius.jl to optimize complex reaction models with high accuracy, good generalization capability, and high efficiency. Such optimization capability will help augment current mechanism reduction techniques.

\subsection{Application to Data Assimilation} \label{sec:datafusion}
While the uncertainty in chemical models is still a standing problem in the combustion community, a general consensus on the major pathways for small hydrocarbons (C$_0$-C$_4$) has been reached. Therefore, such reaction models can be assimilated with experimental data to estimate hidden information from measured quantities \cite{raissi2020hidden} and guide further model refinements.

 Here, we applied Arrhenius.jl to the pyrolysis of propane in shock tubes to assimilate the measured concentration-time series data for all major species with the detailed kinetic mechanisms of USCMech II \cite{wang2007usc}. Again, we relied on Arrhenius.jl to efficiently compute the loss functions to 333 Arrhenius parameters in the propane pyrolysis model. The results demonstrate the capability of Arrhenius.jl in data-assimilation for inferring the temperature profiles from species profiles and suggesting changes of rate constants.

\begin{figure}[!h]
    \centering
    \includegraphics[width=1.0\linewidth]{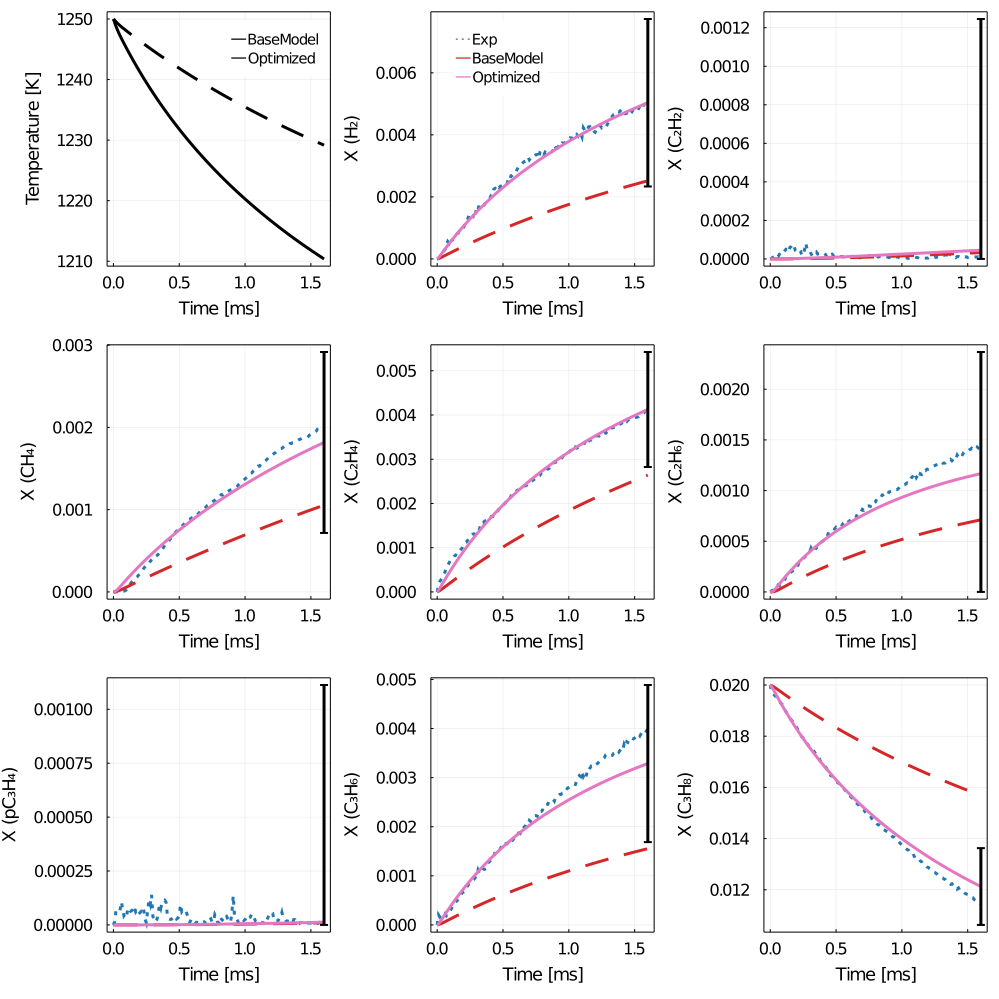}
    \caption{Training dataset: comparisons among measurements, predictions from the baseline model, and predictions from the optimized model. The error-bar corresponds to the estimated experimental uncertainties. Details on the experimental procedure and estimation of experimental uncertainties can be found in \cite{cassady2020pyrolysis}.}
    \label{fig:propane_1}
\end{figure}

\begin{figure}[!h]
    \centering
    \includegraphics[width=1.0\linewidth]{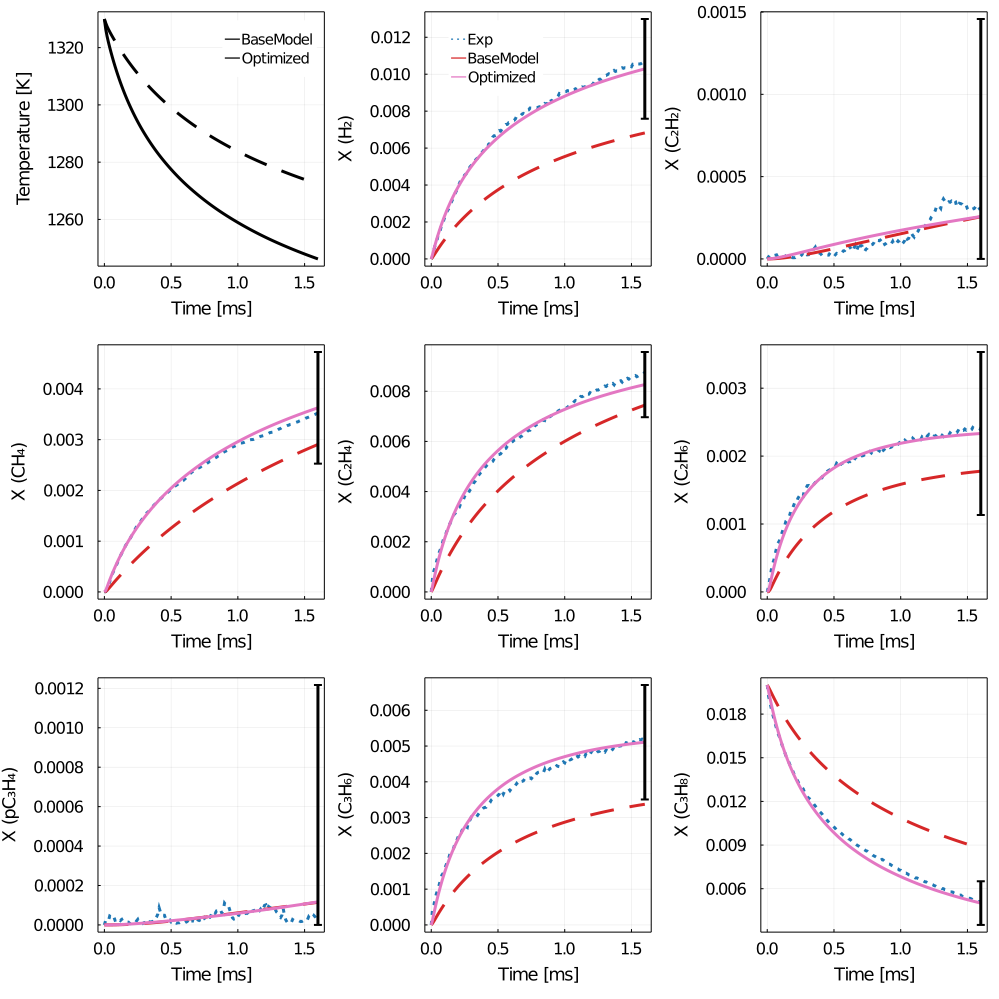}
    \caption{Validation dataset: comparisons among measurements, predictions from the baseline model, and predictions from the optimized model.}
    \label{fig:propane_3}
\end{figure}

The pyrolysis of propane plays a crucial role in characterizing the combustion behaviors of natural gas mixtures and provides insight into the cracking pathways of larger fuels. We adopted the recently available experimental dataset measured in \cite{cassady2020pyrolysis} using a novel laser absorption technique based on convex optimization. The dataset includes eight species profiles shown in Figs. \ref{fig:propane_1} and \ref{fig:propane_3}. The data includes five initial conditions, which are at 4 atm, 1250-1370 K, and with 2\% propane in argon. USCMech II was adopted as the baseline kinetic model. Similar to the comparisons in \cite{cassady2020pyrolysis} to AramcoMech 3.0 \cite{zhou2018experimental}, current kinetic models systematically over-predict the mole fractions of propane under all temperatures. Therefore, we optimized the 111 reactions that are directly related to the pyrolysis of propane. The simulations were carried out under constant pressure, and the governing equations for the mass fractions of species and temperature were solved. Similar to the studies of deep reduction, we optimized all three Arrhenius parameters of these 111 reactions. The parameters were also scaled according Eq. \ref{eq:pdef} but with the unit of $Ea$ specified as $kcal/mol$.

The loss function was defined as the mean absolute error (MAE) between the predicted and the measured species profiles, as shown in Eq.~\ref{eq:propane_loss}:
\begin{equation}
    Loss = MAE\left(X^{pred}, X^{exp} \right),
\label{eq:propane_loss}
\end{equation}
where $X^{pred}$ and $X^{exp}$ correspond to the predicted and measured mole fractions of species, respectively.

The five datasets are divided into four training sets and one validation set. The case with the initial temperature of 1330 K, being the median of the five datasets, was chosen as the validation set. Figure \ref{eq:loss} presents the history of loss functions, $L_2$-norm of gradients, and $L_2$-norm of model parameters. While the training loss slightly increases after 500 epochs, we let the training go to 1000 epochs to regularize the changes of parameters. The training took four hours for 1000 epochs (15 seconds/epoch).

\begin{figure}[!h]
    \centering
    \includegraphics[width=1.0\linewidth]{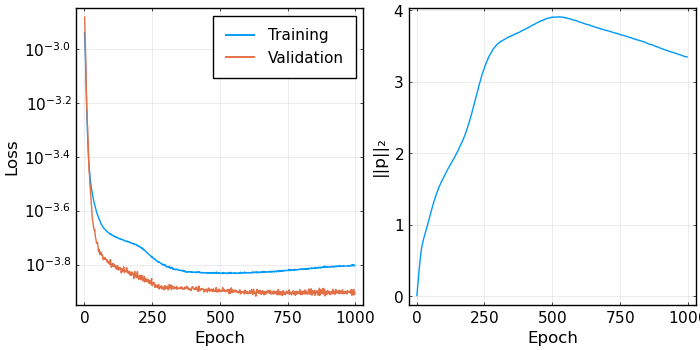}
    \caption{The history of loss functions, $L_2$-norm of model parameters. Regularization settings: weight-decay of 1e-3 with a learning rate of 1e-3.}
    \label{fig:propane_loss}
\end{figure}

The predictions from the optimized model are also shown in Figs. \ref{fig:propane_1} and \ref{fig:propane_3}. The optimized model performs substantially better than the baseline model and agrees very well with the measurements. Since the temperature profiles are linked to the fuel decomposition rates, the predicted temperature profiles are also different from those predicted by the baseline model. For example, the temperature difference is as large as 20 K at 1.5 ms. Furthermore, the optimized mechanism also performs very well with respect to the validation dataset, suggesting that there is no obvious over-fitting in terms of the predictability.

Given the goodness of fit when compared to the measured species profiles, one can expect that the predicted temperature profiles will also agree with the actual temperature time-histories very well. The inferred temperature using the optimized model is qualitatively consistent with the trends in propane concentration as well. For instance, since the pyrolysis process is endothermic, the faster the pyrolysis progresses, the faster the temperature decreases. Therefore, such a data assimilation process could be utilized to build a virtual temperature sensor based on species profiles. In addition, one can utilize the same procedure to infer unknown species concentrations from the available species measurements.

The data assimilation could also be targeted for suggesting key reactions that potentially account for the discrepancies between modeling and measurements. Here, we inspected reactions that have changed their rate constants most during the optimization. To facilitate the comparisons, we lumped the changes of the three Arrhenius parameters into the changes of pre-factor $A$ only, according to Eq. \ref{eq:pdef}. As the changes in rate constants varies with temperature, we present the changes at the reference temperature of 1300 K. The complete list of reactions and the changes of kinetic parameters are provided in the supplemental materials Table S1.

It is found that most of the changes in rate constants are no more than one order of magnitude, while changes to 18 out of 111 reactions are greater than an order of magnitude. Future experimental and theoretical studies could be directed to those 18 reactions. It is worth noting that the large changes in rate constants are not necessarily indicating large uncertainties/errors in the rate constants. The large changes could also be attributed to the randomness in the optimization. In an effort to identify reaction rates that might be good candidates for further study, one may further take the advantage of Arrhenius.jl's differential programming to exploit the Bayesian approach \cite{dandekar2020bayesian}, such as Stochastic Langevin Gradient Descent. Similar to the bound-to-bound approach \cite{frenklach1992optimization}, prior information on these rate constants from theoretical calculations and direct experimental measurements can be readily incorporated into the Bayesian framework. Efficient Bayesian inference will also enable Bayesian experimental design \cite{huan2013simulation} to better allocate experimental resources for the determination of rate constants. While this paper focuses on demonstrating the capability of data assimilation, we shall exploit the Bayesian inference in future works.

In summary, the differential programming in Arrhenius.jl offers us the bandwidth to simultaneously optimize hundreds of model parameters for data assimilation, and such capacity could provide new insights and tools for kinetic model optimization and usage. For example, in the absence of direct measurement, this data assimilation approach could be used to accurately infer system temperature from species measurements, or could be used to identify key reaction rates for which further experimental measurement or modeling work is needed.  Future work will include experiments to validate the ability of the data assimilation approach to accurately infer temperature, through direct temperature measurement, and make use of the Bayesian approach to identify specific reaction rates for future study.




\subsection{Application to Model Discovery}
Finally, we present the application to Scientific Machine Learning (SciML) \cite{rackauckas2020universal} by using Arrhenius.jl to develop a neural-network-based pyrolysis submodel within the HyChem model framework \cite{tao2018physics}. Our recently developed Chemical Reaction Neural Network (CRNN) \cite{ji2021autonomous} approach was employed to develop the neural network model for its interpretability, such that the learned model complies with fundamental physical laws and provides chemical insights, as well as its compatibility with CHEMKIN/Cantera packages. The conventional HyChem-based pyrolysis submodels require expert knowledge on the chemical kinetics which takes years to develop. On the contrary, the CRNN approach aims to autonomously discover the reaction pathways and kinetic parameters simultaneously to accelerate high-fidelity chemical model development. In the following demonstration, the CRNN-HyChem approach was utilized to model the jet fuel of JP10.

\begin{figure}[H]
    \centering
    \includegraphics[width=0.8\textwidth]{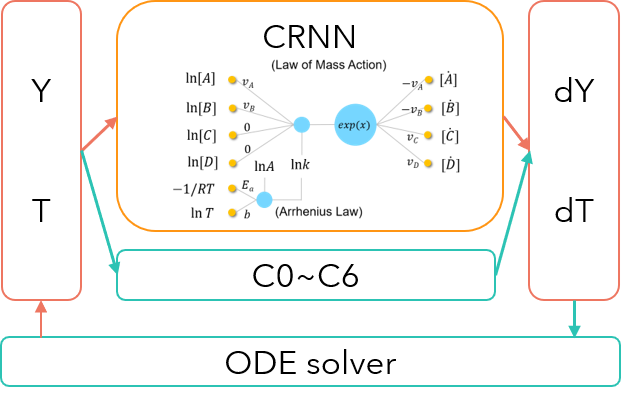}
    \caption{Schematic showing the structure of the CRNN-HyChem approach.}
    \label{fig:schem_crnn_hychem}
\end{figure}

As shown in Fig. \ref{fig:schem_crnn_hychem}, the CRNN-HyChem approach models the fuel chemistry of JP10 with two submodels, similar to the original HyChem concept. The CRNN submodel models the breakdown of JP10 fuel molecules into smaller hydrocarbons up to C$_6$H$_6$, and the submodel for C$_0$-C$_6$ describes the oxidation chemistry. For proof-of-concept, we chose the same species in the original HyChem pyrolysis submodel \cite{tao2018physics} to be included in the CRNN model; however, it should be noted that the such chosen species can be treated as hyper-parameters to circumvent the need for expert knowledge and achieve potentially better performance. In the original CRNN approach \cite{ji2021autonomous}, the Law of Mass Action and Arrhenius Law are enforced by the design of the structure of the neural network. Reaction orders are assumed to be equal to the stoichiometric coefficients for the reactants. In the present study, elemental conservation is further guaranteed by projecting the stoichiometric coefficients into the elemental conservation space. For better convergence, the stoichiometric coefficients for JP10 are fixed as -1, and during the training, the stoichiometric coefficients are regularized to achieve better numerical stability. The training data were generated by simulating the IDT using the original JP10 HyChem model. A wide range of thermodynamic conditions were considered: pressures of 1-60 atm, initial temperatures of 1100-1800 K, and equivalence ratios of 0.5-1.5. In total, 500 thermodynamic conditions were randomly generated using the latin hypercube sampling method. The dataset was split into training and validation datasets with a ratio of 70:30, respectively.

\begin{figure}[H]
    \centering
    \includegraphics[width=1.0\textwidth]{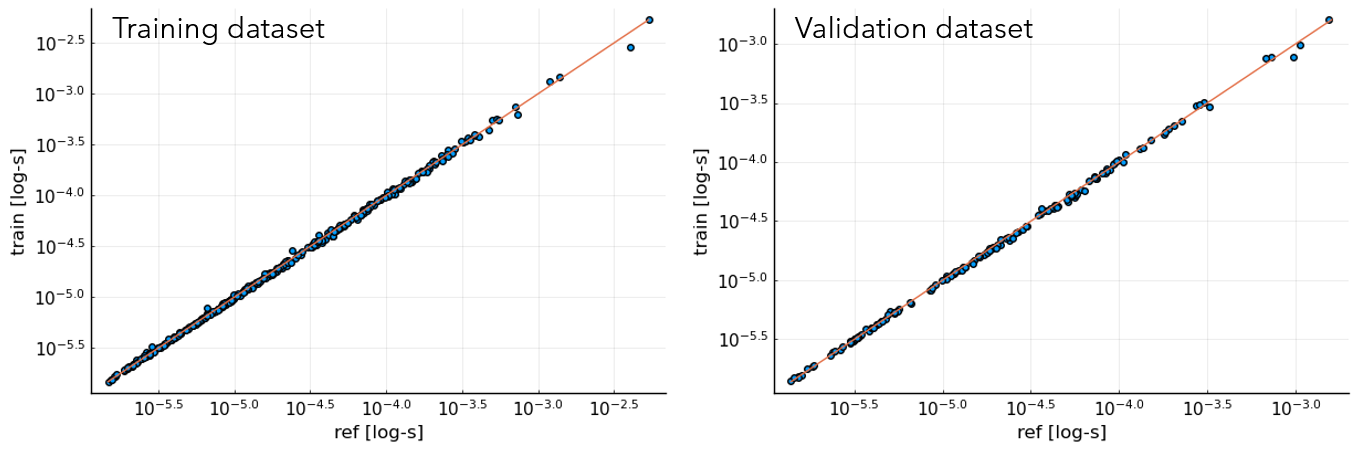}
    \caption{Comparisons between the predicted ignition delay times using the learned CRNN model (Y-axis) and the HyChem model (X-axis) for both the training and validation datasets.}
    \label{fig:regression}
\end{figure}

The learned stoichiometric coefficients and kinetic parameters are shown in Table \ref{table:1}, where negative and positive stoichiometric coefficients correspond to reactants and products, respectively. Qualitatively, most of the learned pathways are H-abstraction reactions, which is consistent with the expert-derived HyChem models. However, quantitatively, the learned pathways are not the same as those in the HyChem model, and further efforts will be directed to extracting physical insights from the learned pathways. Figure \ref{fig:regression} compares the results of the learned CRNN model and the IDTs generated using the original HyChem model \cite{tao2018physics}, and they agree very well. The results thus demonstrate the capability of Arrhenius.jl in learning CRNN models with hundreds of parameters, which is impossible with finite difference methods. With the increasing demand for rapidly developed kinetic models for new renewable fuels for screening and fuel design, CRNN provides an elegant approach to autonomously derive kinetic models from experimental data. Arrhenius.jl will play a vital role in enabling such kind of autonomous model discovery algorithms. While this manuscript focuses on learning the reaction pathways from scratch, one can also develop a data-driven model based on existing kinetic models for similar fuels and utilize Arrhenius.jl for the training, as demonstrated in \cite{ji2021machine}.

\begin{table}[H]
\centering
\caption{Learned stoichiometric coefficients and kinetic parameters. Reaction orders are assumed to be equal to the stoichiometric coefficients for reactants.}
\begin{tabular}{c}
 \includegraphics[width=1.0\linewidth]{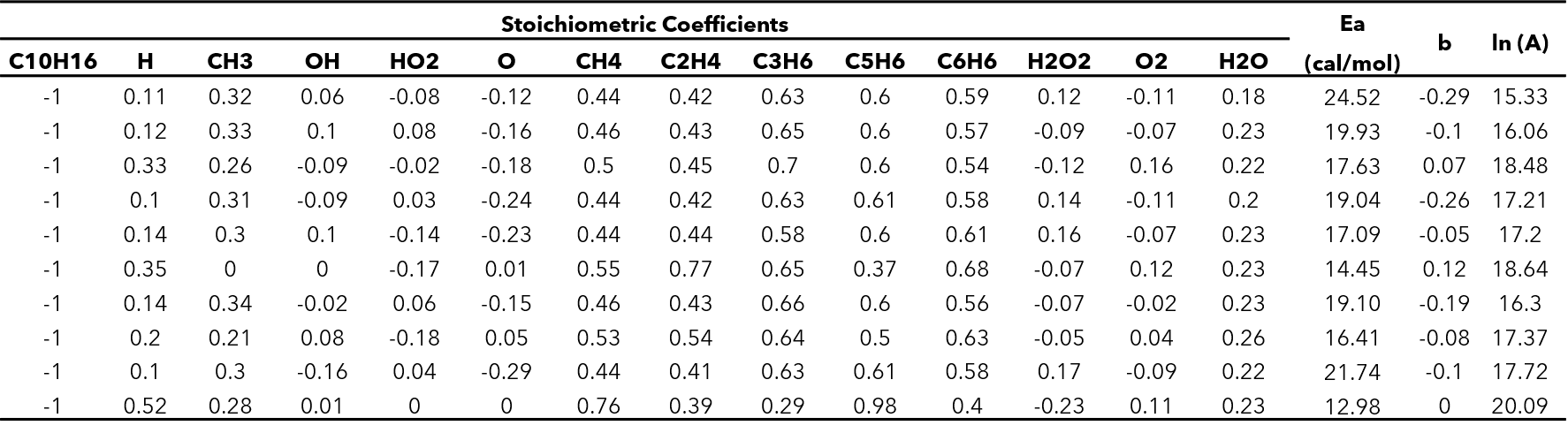}
\end{tabular}
\label{table:1}
\end{table}

\section{Conclusions} \label{sec:conclusion}
This work presents a differential combustion simulation package, Arrhenius.jl, which can perform efficient and accurate gradient evaluations across classical reacting flow solvers. We expect that the open-source package could greatly facilitate the integration of modern deep learning techniques into combustion modeling, especially the physics-informed machine learning that takes the advantage of both physics-based and data-driven modeling. We also invite the contribution from the reacting flow community to enhance the capability of the package and explore its potential in other applications.

\section{Acknowledgments}
WJ and SD would like to acknowledge the funding support by Weichai Power Co., Ltd. ZR and XS would like to acknowledge the support from National Natural Science Foundation of China No. 52025062. WJ would like to thank Dr. Ji-Woong Park for fruitful discussions on the deep mechanism reduction, Dr. Vyaas Gururajan on the implementation of the sensBVP method, Dr. Travis Sikes for discussions on mechanism optimization, Dr. Christopher Rackauckas on the usages of DifferentialEquations.jl, and Matthew Johnson on sharing experience of developing ReactionMechanismSimulator.jl.

\bibliographystyle{CNF} 
\bibliography{ms_betterbib}

\end{document}